\documentclass[aps,prl,twocolumn,showpacs,floatfix]{revtex4-1}
\usepackage{amsmath}    
\usepackage{graphicx}   
\usepackage{verbatim}   
\usepackage{color}      
\usepackage{subfigure}  
\usepackage{hyperref}   

\usepackage{amsmath}

\graphicspath{{./Figures/}}
\hypersetup{pdfborder = {0 0 0},colorlinks=true}

\newcommand{\s}{s} 
\newcommand{\sss}{\pmb{s}} 

\newcommand{\J}{\boldsymbol{J}} 
\newcommand{\h}{\mathbf{h}} 
\newcommand{\ddh}{\delta\mathbf{h}} 

\begin{document}
\title{Network inference in the non-equilibrium steady state}

\author{Simon L. Dettmer$^1$, H. Chau Nguyen$^2$ and Johannes Berg$^1$}
\email{sdettmer@thp.uni-koeln.de, chau@pks.mpg.de, and berg@thp.uni-koeln.de}
\affiliation{$^1$Institute for Theoretical Physics, University of Cologne,
  Z\"ulpicher Stra{\ss}e 77, 50937 Cologne, Germany \\
$^2$Max-Planck-Institut f\"ur Physik komplexer Systeme,
N\"othnitzer Str. 38, 01187 Dresden, Germany}

\begin{abstract} Non-equilibrium systems lack an explicit characterisation of their steady state like the Boltzmann distribution for equilibrium systems. This has drastic consequences for the inference of parameters of a model when its dynamics lacks detailed balance. Such non-equilibrium systems occur naturally in applications like neural networks or gene regulatory networks. Here, we focus on the paradigmatic asymmetric Ising model and show that we can learn its parameters from independent samples of the non-equilibrium steady state. We present both an exact inference algorithm and a computationally more efficient, approximate algorithm for weak interactions based on a systematic expansion around mean-field theory. Obtaining expressions for magnetisations, two- and three-point spin correlations, we establish that these observables are sufficient to infer the model parameters. Further, we discuss the symmetries characterising the different orders of the expansion around the mean field and show how different types of dynamics can be distinguished on the basis of samples from the non-equilibrium steady state.  \end{abstract}



\maketitle

Inverse problems in statistical physics are motivated by the challenges of big data in different fields, especially high-throughput experiments in biology: Can one learn, for instance,  the synaptic connections between neurons from observations of neural activity~\cite{Cocco2009a}, or the regulatory interactions between genes from gene expression levels~\cite{Dhaeseleer2000a}? Inverse statistical problems such as these also include the determination of three-dimensional protein structures~\cite{Weigt2009a}, the inference
of fitness landscapes~\cite{diverseFitness}, and flocking dynamics~\cite{bialek2012a}. The paradigmatic inverse statistical problem is the inverse Ising problem, which seeks to infer the parameters of a spin model -- external fields and interactions between spins -- from observables like magnetisations and spin correlations. For the equilibrium case, which is characterised by symmetric interactions between pairs of spins and detailed balance, a wide range of approaches have been developed on the basis of the Boltzmann distribution~\cite{diverseInvIsing}. Yet in many applications, like neural networks or gene regulatory networks, interactions are generally asymmetric. For instance, a synaptic connection from neuron $A$ to neuron $B$ does not imply a reverse connection back from $B$ to $A$. The resulting non-equilibrium steady state (NESS) lacks detailed balance and is not described by a Boltzmann distribution. We ask if it is possible to learn the parameters of a non-equilibrium model from independent samples of the NESS, even though the probability with which each state of the system occurs in the long-time limit is unknown. Key result of this paper is the identification of observables from which the parameters of a non-equilibrium model can be reconstructed, and a systematic procedure to infer the model parameters based on these observables.  

For concreteness, we start with the asymmetric Ising model in discrete time under so-called Glauber dynamics~\cite{glauber1963a}: 
At a given time $t$, the state of a system of $N$ binary spins is characterised by variables $s_i(t)$ with $i=1,\ldots,N$. The dynamics of spins is defined by randomly picking a spin variable, say $i$, at each time step. The value of that spin variable is then updated, with $\s_i(t+1)=\pm 1$ sampled from the probability distribution
\begin{equation}
\label{eq:glaubersq}
w(\s_i(t+1)|\sss(t))=\frac{\exp\{\s_i(t+1) \theta_i(t)\}}{2
  \cosh(\theta_i(t))} \ ,
\end{equation}
where the effective local field is 
\begin{equation}
\label{eq:localfield}
\theta_i(t)=h_i+\sum_{j=1}^N J_{ij} \s_j(t)  \ ,
\end{equation}
with external fields $h_i$ and couplings between spins $J_{ij}$.
This dynamics has been used as a model of neural dynamics~\cite{Derrida1987a} and as a model of gene expression dynamics~\cite{Bailly2010b}. 
Similar to this sequential dynamics that updates one spin after the other, one can define a dynamics with parallel updates, where the updates~\eqref{eq:glaubersq} are carried out simultaneously for all spins.

For a symmetric coupling matrix without self-couplings, the Glauber dynamics~\eqref{eq:glaubersq} converges to the equilibrium state characterised by the Boltzmann distribution ${p_B(\mathbf{s})=e^{-\mathcal{H}(\mathbf{s})}/Z}$ with the well-known Ising Hamiltonian ${\mathcal{H}(\mathbf{s})=-\sum_i s_i (h_i+\sum_{j>i} J_{ij}s_j)}$.  (For convenience we have subsumed the inverse temperature into the couplings and fields.) For asymmetric couplings, however, Glauber dynamics~\eqref{eq:glaubersq} converges to a non-equilibrium steady state, which lacks detailed balance and is hard to characterise.

In the inverse problem, the task is to infer the parameters of the asymmetric Ising model, namely the couplings $J_{ij}$ and external fields $h_i$ of \eqref{eq:localfield}. This is a comparatively easy task when we can observe time series of consecutive states of the system $\sss(t),\sss(t+1),\sss(t+2),\ldots$. Using the dynamical rule~\eqref{eq:glaubersq}, the probability of a particular trajectory $\prod_t w(\sss(t+1)|\sss(t))$ can be written down explicitly and be maximised with respect to the couplings and fields~\cite{Roudi2011a,zeng2013} in polynomial time in $N$ and the length of the trajectory. This yields the maximum likelihood estimate of the model parameters. An estimate of the couplings and fields that can be computed even faster has been derived using mean-field theory~\cite{Roudi2011a,Mezard2011a}.

However, there are situations where a time series of the system's dynamics is not available. An example is genome-wide gene expression levels measured in single cells, a process which involves the physical destruction of cells. 
In such cases, only independent samples from the non-equilibrium steady state are available. Beyond the practical relevance, it is also a fundamental question whether we can characterise the NESS sufficiently well to solve the inverse problem.

Already elementary arguments show that, unlike in the equilibrium case~\cite{Ackley1985a,diverseInvIsing}, pairwise spin-correlations are insufficient to infer the model parameters: the matrix of 
pairwise correlations $\langle s_i s_j\rangle$ is symmetric and has only $N(N-1)/2$ independent entries, whereas there are $N(N-1)$ entries of the 
asymmetric coupling matrix $J_{ij}$ to be determined (self-interactions $J_{ii}\neq 0$ are excluded). Thus we expect that at least three-point correlations $\langle s_i s_j s_k\rangle$ are required. On the other hand, the information one can extract from single-time measurements in the NESS is limited to the frequencies of the $2^N$ different spin configurations. Taking into account the normalisation constraint, there are thus at most $2^N-1$ independent observables available to determine the $N(N-1)+N$ parameters of couplings and external fields. This implies that the parameters can only be inferred for $N\geq 5$. 

\paragraph{Self-consistent magnetisations and spin-correlations.} 
In the following, we write the magnetisations and $n$-point spin-correlations self-consistently as single-time expectations in the steady state involving the effective local fields~\eqref{eq:localfield}. Further, we employ an expansion around a probability distribution factorising in the spins (mean-field theory) to derive magnetisations, two-, and three-point correlations as an explicit function of the couplings and external fields. 
By inverting either of these relationships we can solve the inverse problem.

We consider the magnetisations $m_i\equiv \langle s_i \rangle$ in the steady state, and the fluctuations $\delta s_i\equiv (s_i-m_i)$  of spins around this mean. By averaging over the statistics~\eqref{eq:glaubersq}, we obtain for the magnetisations and the $n$-point connected correlations 
$C_{i_1,i_2,\hdots,i_n} \equiv \langle \delta s_{i_1} \delta s_{i_2}\cdots \delta s_{i_n}\rangle$ 
\begin{equation}
m_i=\langle \tanh(\theta_i) \rangle \equiv \sum_{\sss} p(\sss|\h,\J) \tanh(\theta_i(\sss,\h,\J)) 
\label{eq:self-consistent-magnetisations}
\end{equation}
\begin{equation}
C_{i_1,\hdots,i_n}= \frac{1}{n}\sum_{k=1}^n \left\langle \left(\prod_{\substack{j=1\\ j\neq k}}^n\delta s_{i_j}\right)[\tanh(\theta_{i_k})-m_{i_k}] \right \rangle ,
\label{eq:self-consistent-correlations}
\end{equation}
where $\{i_1,\hdots,i_n\}\subset \{1,\hdots N\}$ is a subset of $n$ spins and $p(\sss|\h,\J)$ denotes the steady-state probability over spin configurations $\sss$. 
These equations are a set of self-consistent equations whose left-hand side gives magnetisations and correlations, which together specify the probability distribution $p(\sss|\h,\J)$ in the the NESS. The right hand sides depend on this distribution via 
averages over functions of $\theta_i = h_i+\sum_{j=1}^N J_{ij} \s_j$. 
A similar result holds for a dynamics with parallel updates $
{C_{i_1,\hdots,i_n}^\text{par}=\left \langle \prod_{k=1}^n[\tanh(\theta_{i_k})-m_{i_k}] \right  \rangle }$.
Given independent samples from the steady state, the expectations in \eqref{eq:self-consistent-magnetisations} and \eqref{eq:self-consistent-correlations} can be evaluated numerically by averaging over the sampled configurations.

\paragraph{An expansion around mean-field theory.}
Although the steady-state probability distribution $p(\sss|\h,\J)$ underlying the expectations is not known, it turns out that its moments can be calculated in a systematic expansion around a distribution factorising in the spins, ${q(\sss|\mathbf{h}_q) \equiv p(\sss|\mathbf{h}_q,\J_q=0)}$. In equilibrium statistical physics, this distribution is the well-known mean-field ansatz. Its application to the non-equilibrium setting has been pioneered by Kappen and Spanjers~\cite{Kappen2000a}. As usual in mean-field theory, the external fields ${\mathbf{h}_q=\mathbf{h}_q(\mathbf{h},\J):=\mathbf{h}-\delta \mathbf{h}}$ characterising the mean-field distribution $q(\sss|\mathbf{h}_q)$ are fixed by a self-consistent equation for the magnetisations ${\mathbf{m}(\mathbf{h}_q,\J_q=0)=\mathbf{m}(\mathbf{h},\J)}$, so the mean-field distribution yields the same magnetisations as the original model with couplings $\J$ and fields $\h$.  Considering external fields $\mathbf{h}_q+\lambda\ddh $ and couplings $\lambda \J$, one can smoothly interpolate between $\lambda=0$, describing the factorising distribution $q$, to the NESS described by $\lambda=1$.  Expanding the moments of this distribution in a Taylor-series around the mean-field distribution $\lambda=0$ gives 
\begin{eqnarray} 
\label{eq:kappenspanjers} 
\mathbf{m}(\mathbf{h},\J)&=&\sum_{k =0}^{\infty} \frac{1}{k!} \frac{\text{d}^k \mathbf{m}(\mathbf{h}_q+\lambda\ddh,\lambda \J)}{\text{d} \lambda^k}\Big|_{\lambda=0} \\
 \mathbf{C}(\mathbf{h},\J)&=&\sum_{k=0}^{\infty} \frac{1}{k!}\frac{\text{d}^k \mathbf{C}(\mathbf{h}_q+\lambda \ddh,\lambda \J)}{\text{d} \lambda^k}\Big|_{\lambda=0}  \ . \nonumber 
\end{eqnarray}

Using this approach, Kappen and Spanjers computed the magnetisations and two-point correlations to second order in $\lambda$~\cite{Kappen2000a} 
\begin{eqnarray}
m_i&=&\tanh \left(h_i+\sum_{j=1}^N J_{ij}m_j -m_i\sum_{j=1}^N J_{ij}^2(1-m_j^2)\right) \label{eq:TAP} \\
C_{ij}&=&(1-m_i^2)(1-m_j^2)\Bigg(J_{ij}^\text{sym}+ m_im_j \left(J_{ij}^2+J_{ji}^2\right) \notag \\ &&+\sum_{\substack{k=1\\k\neq i}}^N\frac{ J_{jk}J_{ik}^{sym}+J_{ik}J_{jk}^{sym}
}{2} (1-m_k^2)  \Bigg)\label{eq:two-point-TAP}  \ ,
\end{eqnarray}

where $\J^\text{sym}=\frac{1}{2}(\J+\J^T)$ and $\J^\text{asym}=\frac{1}{2}(\J-\J^T)$ are the symmetric and antisymmetric parts of the coupling matrix respectively.
\paragraph{Three-point correlations and their symmetries.} To infer couplings and fields we turn to the connected three-point correlations. To second order in $\lambda$ we find 
\begin{eqnarray}
\label{eq:three-point-TAP}
 C_{ijk}=&\frac{1}{3}(1-m_i^2)(1-m_j^2)(1-m_k^2)\times \\ &[-6A_{ijk}(\J^\text{sym},\mathbf{m})-2A_{ijk}(\J^\text{asym},\mathbf{m})] \ , \nonumber
\end{eqnarray}
where
\begin{equation}
A_{ijk}(\J,\textbf{m})=J_{ij} J_{kj}m_j+ J_{ji} J_{ki}m_i+ J_{jk}J_{ik}m_k \ .
\end{equation}

These spin correlations exhibit particular symmetries, which affect the reconstruction of model parameters. Already the two-point correlations~\eqref{eq:two-point-TAP} depend, to first order in $\J$, only on the symmetric part of the coupling matrix. However, also the three-point correlations show a symmetry;~\eqref{eq:three-point-TAP} is unchanged when the coupling matrix is replaced with its transpose so $\J^\text{asym}$ transforms to $-\J^\text{asym}$, since $A_{ijk}(\J,\textbf{m})$ is quadratic in the couplings. Thus jointly solving~\eqref{eq:two-point-TAP} and \eqref{eq:three-point-TAP} for the coupling matrix $\J$ either yields the reconstruction of the original coupling matrix, or its transpose. This binary symmetry is lifted only at third order in $\lambda$, see Supplemental Material.

\paragraph{Parameter inference.} 

Given empirical samples from the NESS we can now solve the inverse problem in two ways: (i) \textit{exact inference.} We jointly solve the self-consistent equations \eqref{eq:self-consistent-magnetisations} and \eqref{eq:self-consistent-correlations} up to three-point correlations for the couplings $\J$ and external fields $\mathbf{h}$.
(ii) \textit{mean-field inference.} We jointly solve the explicit correlation expressions~\eqref{eq:two-point-TAP} and~\eqref{eq:three-point-TAP} (taken to third order in $\lambda$) for the coupling matrix $\J$. Subsequently solving the magnetisation equations~\eqref{eq:TAP} for $\mathbf{h}$ completes the parameter reconstruction.

To test these inference schemes, we numerically simulate a system of $N=10$ spins with random asymmetric couplings. Off-diagonal entries of the matrix of couplings are chosen independently from a Gaussian distribution with zero mean and standard deviation $\beta/\sqrt{N}$ (self-interactions are excluded: $J_{ii}\equiv0$) , and external fields from a Gaussian distribution with zero mean and standard deviation $\beta$. Samples of the state of the system under Glauber dynamics \eqref{eq:glaubersq} with sequential updates are recorded at each update after an initial settling-in period of $10^5 N$ updates to reach the steady state. 
Based on these measurements, we reconstruct the parameters by minimising the sum of the relative squared prediction errors of the magnetisation and two- and three-point correlation equations using the Levenberg-Marquardt algorithm (see Supplemental Material for details). 

Figure~\ref{fig1} shows the reconstruction of the couplings for different numbers of samples and coupling strengths. Three-point correlations are small and as a result the inference is affected by sampling noise. For the exact inference, the reconstruction improves significantly with the number of samples (left hand plots). For the mean-field inference, the correlations~\eqref{eq:two-point-TAP}-\eqref{eq:three-point-TAP} computed to finite order in $\lambda$ become inaccurate in the limit of strong couplings, which can also limit the reconstruction quality. As a result, the mean-field reconstruction performs best for intermediate coupling strengths (right hand plots). Also, the reconstruction error for the symmetric part of the couplings $\J^\text{sym}$ is smaller than for the antisymmetric part $\J^\text{asym}$, since the former is primarily determined by the connected two-point correlations~\eqref{eq:two-point-TAP}, which are considerably larger than the three-point correlations. For this reason, fewer samples are required for the accurate inference of the symmetric part of the couplings. The reconstruction of the external fields shows a similar behavior; see Fig.~S1 in the Supplemental Material.

\begin{figure}[tbh!]
\includegraphics[width = .5\textwidth]{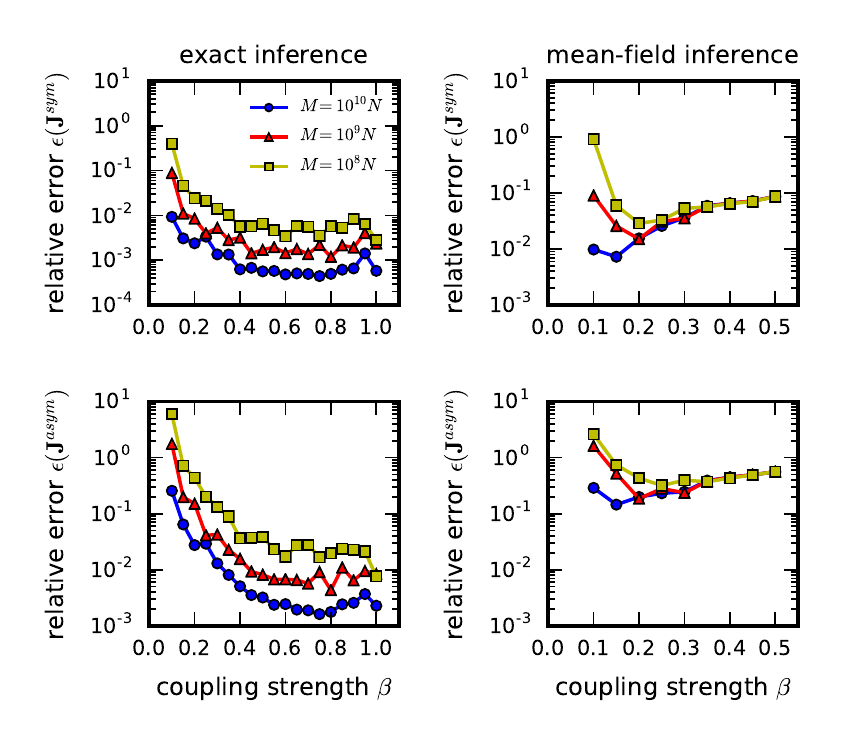}
\caption{\textbf{Couplings inferred from the NESS}. We consider a system of $N=10$ spins with random asymmetric couplings under Glauber dynamics~\eqref{eq:glaubersq}, see text. We plot the relative root-mean-squared reconstruction error $\epsilon$ between the inferred and the true
couplings against the coupling strength $\beta$ for different numbers of samples $M$. Left: reconstruction errors for the exact inference. Right: reconstruction errors for the mean-field inference. The symmetric part of the couplings $\J^\text{sym}=(\J+\J^T)/2$ (top) generally has a lower reconstruction error than the antisymmetric part $\J^\text{asym}=(\J-\J^T)/2$ (bottom).
}
\label{fig1}
\end{figure}

\begin{figure}[tbh!]
\center
\includegraphics[width = .45\textwidth]{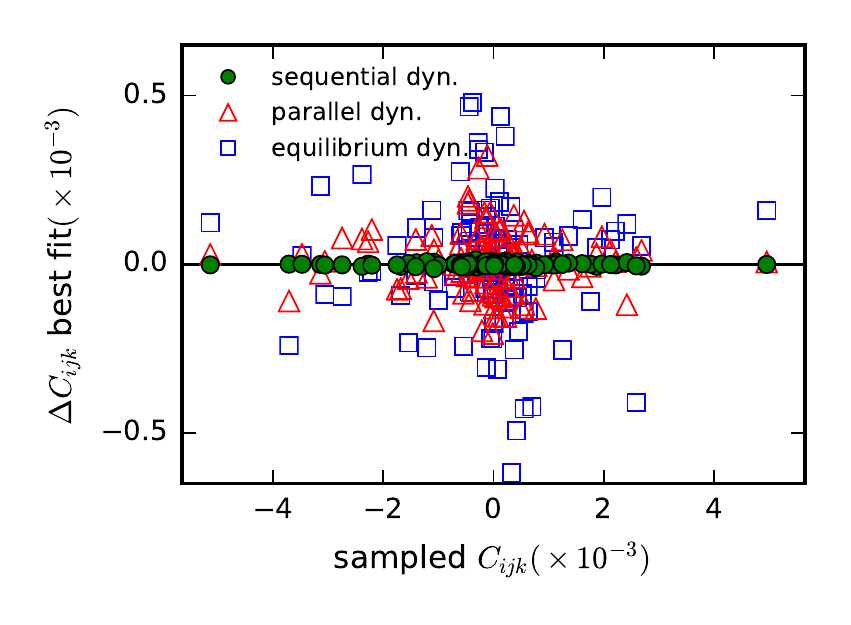}
\caption{\textbf{Distinguishing different dynamical models on the basis of three-point correlations.} We sample configurations from the NESS of Glauber dynamics with sequential updates, see text. 
Next, we reconstruct the model parameters from the magnetisations, two- and three-point correlations, assuming the data was generated by Glauber dynamics with sequential updates (circles), parallel updates (triangles), or equilibrium dynamics (squares).  
The deviations $\Delta C_{ijk}=C_{ijk}(\h,\J)-C_{ijk}^\text{sampled}$ of the three-point correlations predicted by these three models from the corresponding correlations seen in the original samples are plotted against the sampled correlations. 
 The relative root-mean-squared prediction errors $\| \Delta C_{ijk}\|_2/\| C_{ijk}^\text{sampled}\|_2$ are $0.003$,  $0.06$, and $0.13$ for the sequential, parallel, and equilibrium dynamics respectively, clearly favouring the dynamics with sequential updates. The horizontal line is a guide to the eye representing a perfect fit. We used $N=10$ spins, a coupling and external field strength of $\beta=0.2$ and $M=10^{10}N$ samples. } 
\label{fig2}
\end{figure}

\paragraph{Model selection.} Beyond estimating the parameters of a particular dynamical model, an important question is what \emph{type} of dynamics produced a particular NESS. In inference, this question is known as the model selection problem.  Here, we compare three different dynamics: (i) Glauber dynamics with sequential updates, (ii) Glauber dynamics with parallel updates, and (iii) equilibrium dynamics (sequential updates with $\J^\text{asym}=0$). We start by taking independent samples from the NESS produced by a model with sequential dynamics as described above and calculate magnetisations and correlations. Next, we solve the exact self-consistent equations for the magnetisations, two- and three-point correlations for the different dynamics by minimising the relative prediction error as above. This gives the model parameters for a particular dynamics that best reproduce the sampled correlations. In Fig.~\ref{fig2} we compare the three-point correlations predicted by these best fits of the three different dynamical models with the sampled correlations. Indeed, the sequential model shows the best match with the sampled data, leading to the conclusion that out of the three alternatives, the data was indeed most likely produced by a model with sequential Glauber dynamics. We find analogous results for the dynamics generated by parallel updates, see Fig.~S2 in the Supplemental Material. This shows that one can distinguish the different types of dynamics based on samples from their NESS alone.

\paragraph{The inverse Langevin problem.} Our approach is not limited to the asymmetric Ising problem with its binary spins and discrete-time dynamics. Consider the dynamics of continuous variables $x_i$ under a model of the form 
\begin{equation}
\label{eq:langevin}
\partial_t x_i=f(\theta_i)  -x_i+\xi_i(t) \ ,
\end{equation}
where the effective local field is $\theta_i(t)=h_i+\sum_j J_{ij} x_j(t)$, $f(\theta)$ is an arbitrary monotonic function, and $\xi_i(t)$ describes $\delta$-correlated white noise. This is a multivariate Langevin equation; the steady state, if it exists, generally does not obey detailed balance.  For the particular choice $f(\theta)=\tanh(\theta)$, this Langevin equation has magnetisations and correlations given by~\eqref{eq:self-consistent-magnetisations} and~\eqref{eq:self-consistent-correlations}, and these results can be generalised easily to arbitrary choices of $f(\theta)$. Hence these equations and their generalisations can also be used to solve the inverse problem for the class of non-equilibrium stochastic differential equations of the form~\eqref{eq:langevin}.

To conclude, we have used a self-consistent characterisation of the non-equilibrium steady state for the inference of model parameters from independent samples, that is, without direct recourse to the dynamics of the system. We showed that in the case of the asymmetric Ising model, correlations beyond two-point correlation are necessary for parameter inference, and that three-point correlations are in fact sufficient for this task. 

%

\end{document}


\title{Network inference in the non-equilibrium steady state:\\Supplemental Material}

\author{Simon L. Dettmer$^1$, H. Chau Nguyen$^2$ and Johannes Berg$^1$}
\email{sdettmer@thp.uni-koeln.de, chau@pks.mpg.de, and berg@thp.uni-koeln.de}
\affiliation{$^1$Institute for Theoretical Physics, University of Cologne,
  Z\"ulpicher Stra{\ss}e 77, 50937 Cologne, Germany \\
$^2$Max-Planck-Institut f\"ur Physik komplexer Systeme,
N\"othnitzer Str. 38, 01187 Dresden, Germany}

\maketitle

\section{Mean field equations to third order in $\lambda$}
Here we give the results of expanding the magnetisations, two-~and three-point correlations
(5) to third order in $\lambda$. 

\subsection{Magnetisations}

Continuing the expansion (5) for the magnetisations to third order in $\lambda$, we
find 
\begin{eqnarray}
\label{eq:mag_MF_3rd}
m_i=\tanh \Big(&\theta_i+\sum_{j=1}^N J_{ij}m_j -m_i a_i\\
&+\frac{2}{3}(1-3m_i^2)b_i-m_i c_i \Big) \ , \nonumber
\end{eqnarray}
where we defined the auxiliary quantities
\begin{eqnarray}
\label{eq:aux1}
a_i=&\sum_{j=1}^N J_{ij}^2(1-m_j^2) \\
b_i=&\sum_{j=1}^N J_{ij}^3 m_j(1-m_j^2) \\
c_i=&\sum_{\substack{j,k=1 \\ j\neq k}}^N J_{ij}(1-m_j^2)J_{ik}(1-m_k^2)J_{jk}^\text{sym}
\end{eqnarray}
and $J_{ij}^\text{sym}=(J_{ij}+J_{ji})/2$ denotes the entry of the symmetric part of the coupling matrix.
This result differs from the one obtained using the Plefka expansion
of the equilibrium free energy~\cite{Georges1991}, although 
naturally both expressions coincide for the special case of symmetric couplings. 

\subsection{Two- and three-point correlations under sequential
 dynamics}

We denote the two- and three-point correlations computed to second order in $\lambda$ by $C_{ij}^\text{TAP}$ and  $C_{ijk}^\text{TAP}$, respectively. Their expressions were given in the main text. 
%

For the third order correction to the two-point correlations we obtain
\begin{align}
\label{eq:two-point-MF}
C_{ij}-C_{ij}^\text{TAP}=&(1-m_i^2)(1-m_j^2)\times \\
 \Bigg\{&\frac{1}{3}J_{ij}^3(1-3m_i^2)(1-3m_j^2)+2m_im_jJ_{ij}A_{ji}  \notag \\
&-\frac{1}{2}J_{ij}(1-m_i^2)a_i+\frac{1}{2}m_iF_{ij}+\frac{1}{2}E_{ij}
\Bigg\}\notag \\
& +(i \leftrightarrow j), \notag
\end{align}
where we defined the auxiliary quantities
\begin{align}
A_{ij}&=\sum_{\substack{k=1 \\ k\neq i}}^N J_{ik}^\text{sym}J_{jk}(1-m_k^2) \\
E_{ij}&=\sum_{\substack{k=1 \\ k\neq i}}^N \frac{A_{ik}+A_{ki}}{2}J_{jk}(1-m_k^2) \\
F_{ij}&=\sum_{\substack{k=1 \\ k\neq i}}^N
  (J_{ik}^2+J_{ki}^2)J_{jk}(1-m_k^2)m_k \ .
\end{align}
$a_i$ are defined in~\eqref{eq:aux1}.

For the third order correction to the three-point correlations we find
\begin{align}
\label{eq:three-point-MF}
C_{ijk}-C_{ijk}^{TAP}=&\frac{(1-m_i^2)(1-m_j^2)(1-m_k^2)}{3}\times \\
 \Bigg\{&-J_{ij}^\text{sym}\left[ 4J_{ki}J_{kj}m_im_jm_k+4m_k(J_{ki}^2m_i^2)\right] \notag \\
&+2J_{ki}m_i \Big[(1-3m_k^2)J_{ki}J_{kj}-m_im_j(J_{ij}^2+J_{ji}^2) \notag \\
&-A_{ij}\Big]
-2m_k\sum_{\substack{l=1 \\l \neq i,j}}^NJ_{kl}(1-m_l^2)J_{ki}J_{jl}^\text{sym} \notag \\
&+\sum_{\substack{l=1 \\l \neq i,j}}^N \frac{J_{kl}}{2}\left(\frac{C_{ijl}^{TAP}}{(1-m_i^2)(1-m_j^2)} \right)\Bigg\}\notag \\
& +\text{permutations of } (i,j,k)  .\notag
\end{align}

\section{Description of the parameter inference algorithm}
In order to reconstruct the model parameters, we consider equations (3) and (4) for the magnetisations, two- and three-point correlations, and solve for the couplings and external fields. Our goal is to minimise the relative squared error of the predictions which leads us to define the cost function 

\begin{eqnarray}
\label{eq:exact_cost}
E(\h,\J)=&\frac{\| \mathbf{m}^\text{sc}(\h,\J)-\mathbf{m}^\text{sampled}\|_2^2}{\| \mathbf{m}^\text{sampled}\|_2^2} \\
&+\frac{\| \mathbf{C}^\text{sc}_{ij}(\h,\J)-\mathbf{C}_{ij}^\text{sampled}\|_2^2}{\| \mathbf{C}_{ij}^\text{sampled}\|_2^2} \nonumber \\
&+\frac{\| \mathbf{C}^\text{sc}_{ijk}(\h,\J)-\mathbf{C}_{ijk}^\text{sampled}\|_2^2}{\| \mathbf{C}_{ijk}^\text{sampled}\|_2^2}, 
\nonumber
\end{eqnarray}

and its mean-field approximation
\begin{eqnarray}
\label{eq:MF_cost}
E_\text{MF}(\J)=&\frac{\| \mathbf{C}^\text{MF}_{ij}(\J)-\mathbf{C}_{ij}^\text{sampled}\|_2^2}{\| \mathbf{C}_{ij}^\text{sampled}\|_2^2} \\
&+\frac{\| \mathbf{C}^\text{MF}_{ijk}(\J)-\mathbf{C}_{ijk}^\text{sampled}\|_2^2}{\| \mathbf{C}_{ijk}^\text{sampled}\|_2^2}, 
\nonumber
\end{eqnarray}
where the $l_2$-norm $\|.\|_2$ for the symmetric correlation tensors is defined as
${\|\mathbf{X}_{ij}\|_2^2=\sum_{i<j}X_{ij}^2}$ and
${\|\mathbf{X}_{ijk}\|_2^2=\sum_{i<j<k}X_{ijk}^2}$  and the indices sc and MF denote the exact self-consistent equations (3) and (4), and the explicit mean-field expressions (7),\eqref{eq:two-point-MF} and (8),\eqref{eq:three-point-MF} respectively. 

These cost functions are non-negative functions of the model parameters, they are zero when the connected two- and three-point correlations (and magnetisations for the exact inference) exactly
match the empirically measured correlations (and magnetisations). The mean-field cost function only depends on the coupling matrix $\J$ since the reconstruction of the fields
is independent from the coupling inference, and is done by solving the magnetisation equation \eqref{eq:mag_MF_3rd} with the reconstructed couplings. 
The cost functions \eqref{eq:exact_cost} and \eqref{eq:MF_cost} can be viewed  as formal energies to be minimised by some algorithm.

Our inference problem can now be restated as
finding ${(\h^*,\J^*)_\text{exact}=\operatorname{argmin}_{\h,\J} E(\h,\J)}$ for the exact inference, or ${\J_\text{MF}^*=\operatorname{argmin}_{\J} E_\text{MF}(\J)}$, $\h_\text{MF}^*=\textbf{m}^{-1}(\J_\text{MF}^*$) for the mean-field inference.

Since this requires solving a system of nonlinear equations, this cannot be done analytically. We use the Levenberg-Marquardt algorithm~\cite{Levenberg1944,Marquardt1963} as implemented in the Python library SciPy~\cite{SciPy} as a numerical solver.
We find that the energy landscapes exhibit many local minima. For that reason it is not sufficient to use a single starting point for the solver. Instead we use 100 random starting points drawn from the same distribution as the model parameters. For each of these starting points, the Levenberg-Marquardt algorithm finds a local minimum and of these candidates we choose the one with the lowest energy.

\section{Reconstruction of the external fields}
\begin{figure}[t!]
\includegraphics[width = .5\textwidth]{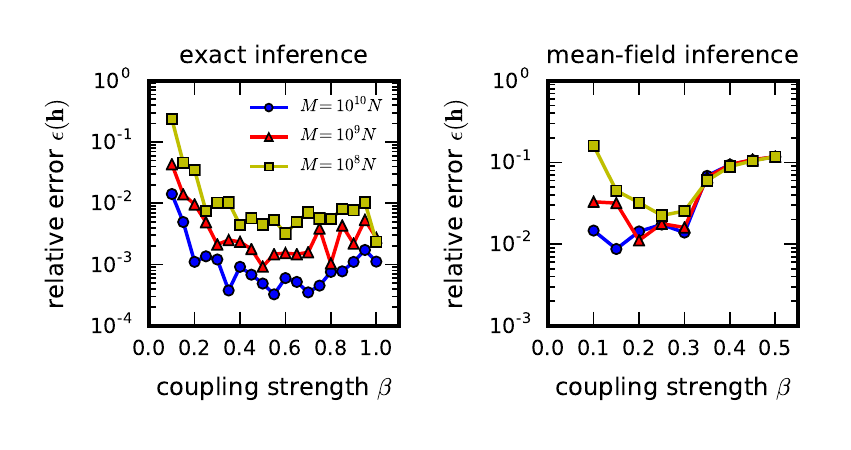}
\caption{\textbf{External fields inferred from the NESS}. This is the
  companion figure to Fig.1 in the main text. We plot the relative root-mean-square
  reconstruction error between the inferred and true external fields, 
${\epsilon(\h)=\|\h^\text{inf}-\h\|_2/\|\h\|_2}$, against the coupling and external field strength $\beta$ for the exact inference (left) and the mean-field inference (right).} 
\label{figS1}
\end{figure}

For the exact inference algorithm, the external fields are inferred simultaneously with the couplings by minimising the exact cost function~\eqref{eq:exact_cost}. Within the mean-field inference scheme, we first reconstruct the couplings from the two- and three-point correlations by minimising the mean-field cost function~\eqref{eq:MF_cost}.
Subsequently, the external fields can be inferred by directly solving the magnetisation equation~\eqref{eq:mag_MF_3rd} for the external fields $\h$.
Results for the coupling inference are shown in Figure~1 in the main text. Figure~\ref{figS1} shows that the reconstruction of the external fields exhibits the same behavior as the reconstruction of the couplings, although at much lower errors. 

\section{Distinguishing different dynamics for data generated with parallel Glauber dynamics}
\begin{figure}[htb!]
\includegraphics[width = .45\textwidth]{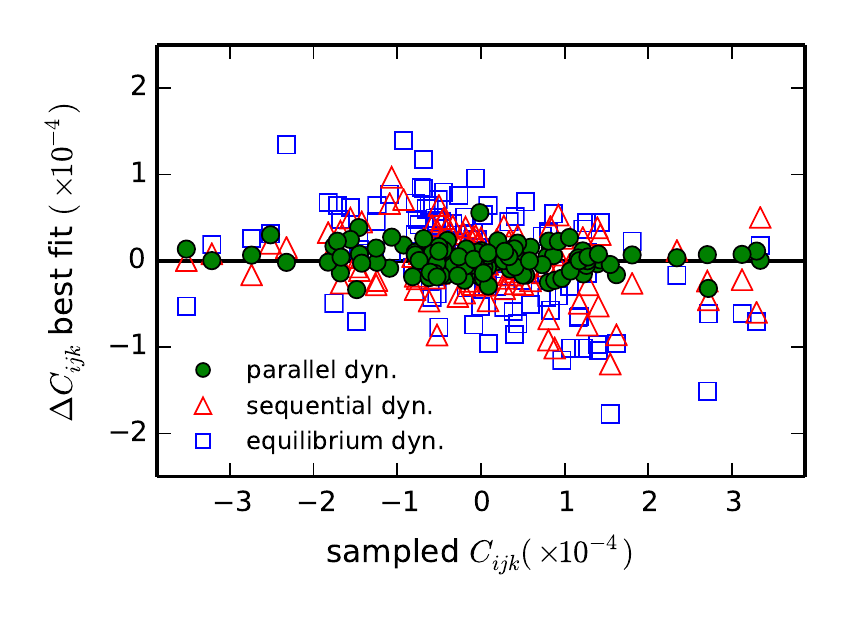}
\caption{\textbf{Detecting parallel dynamics on the basis of three-point
    correlations.} We sample magnetisations and correlations from the
  NESS of Glauber dynamics with parallel updates~\eqref{eq:glauberpar}, as was done in the
  main text for sequential updates (see Fig.~2). Again model parameters are
  obtained under the three alternative models with parallel (circles), sequential (triangles), and equilibrium dynamics (squares). Their predictions for the three-point correlations $C_{ijk}$ are
  compared to those observed in the original data. Shown are the deviations $\Delta C_{ijk}=C_{ijk}(\h,\J)-C_{ijk}^\text{sampled}$ of the predicted correlations from the sampled correlations against the sampled correlations.
 The relative prediction errors $\|
 \Delta C_{ijk}\|_2/\| C_{ijk}^\text{sampled}\|_2$
 are $0.10$,  $0.25$, and $0.41$ for parallel, sequential, and
 equilibrium dynamics respectively, showing that indeed parallel
 dynamics best describes the data. The horizontal line is a guide to
 the eye representing a perfect fit. We used a coupling and external field strength of $\beta=0.2$ and $M=10^{10}N$ samples. } 
\label{figS2}
\end{figure}

In the main text we showed that, for the case of samples generated by
sequential Glauber dynamics, we can correctly identify the type of
dynamics that produced the correlations. Here we show that the same holds true for data generated with parallel Glauber dynamics, where all spins are updated simultaneously and independently during each timestep under the stochastic update rule
\begin{equation}
\label{eq:glauberpar}
w(\sss(t+1)|\sss(t))=\prod_{i=1}^N \frac{\exp\{\s_i(t+1) \theta_i(t)\}}{2
  \cosh(\theta_i(t))} \ .
\end{equation} 
From this dynamical rule we obtain self-consistent equations for the connected correlations in the NESS,
\begin{equation}
C^\text{par}_{i_1,\hdots,i_n}=  \left \langle \prod_{l=1}^n
  [\tanh(\theta_{i_l})-m_{i_l}]  \right \rangle \ .
\label{eq:self-consistent-parallel}
\end{equation}
Again we compare the three scenarios (i) Glauber dynamics with parallel updates, (ii) Glauber dynamics with sequential updates, and (iii) parallel equilibrium dynamics (parallel updates with $\J^\text{asym}=0$). Magnetisations, two- and three-point correlations were sampled from the NESS generated by parallel Glauber dynamics \eqref{eq:glauberpar}. Subsequently, the exact self-consistent equations for the magnetisations, two- and three-point correlations were fitted to the data for the three different scenarios. Figure~\ref{figS2} shows that we can indeed correctly identify the dynamics that produced the data. 

%